\documentclass{article}
\usepackage{spconf,amssymb, amsmath, graphicx, paralist,subfigure,epsfig,cite}
\usepackage{float}
\usepackage[ruled,vlined,linesnumbered]{algorithm2e}
\usepackage[usenames,dvipsnames]{color}
\usepackage{caption}
\usepackage{soul}

\newtheorem{lem}{Lemma}

\def\mb{\mathbf}

\def\mc{\mathcal}

\title{Simultaneous Distributed Estimation and Attack Detection/Isolation in Social Networks: Structural Observability, Kronecker-Product Network, and Chi-Square Detector}
\name{%
\begin{tabular}{@{}c@{}}
	Mohammadreza~Doostmohammadian$^{\dagger \star}$, Themistoklis Charalambous$^\dagger$,~\textit{Senior Member,~IEEE,} \qquad \\
	Miadreza Shafie-khah$^\ast$,~\textit{Senior Member,~IEEE,} Nader Meskin$^\diamond$,~\textit{Senior Member,~IEEE,} \qquad \\ and~Usman~A.~Khan$^\ddagger$,~\textit{Senior Member,~IEEE }
\end{tabular}
\thanks{This work is partially supported by NSF under awards~\#1903972 and~\#1935555. Corresponding email: {\tt\small doost@semnan.ac.ir, mohammadreza.doostmohammadian@aalto.fi}}}

\address{
	$\dagger$ School of Electrical Engineering, Aalto University, Espoo, Finland.
	\\ 
	$^\star$ Faculty of Mechanical Engineering, Semnan University, Semnan, Iran. \\
	$^\ast$ School of Technology and Innovations, University of Vaasa, Vaasa, Finland
	\\
	$^\diamond$ Electrical 	Engineering Department, Qatar University, Doha, Qatar \\
	$^\ddagger$Electrical and Computer Engineering Department, Tufts University, Medford, MA, USA.}

\begin{document}
\maketitle

%
%
%
%
\begin{abstract}
This paper considers distributed estimation of linear systems when the state observations are corrupted with Gaussian noise of unbounded support and under possible random adversarial attacks.
We consider sensors equipped with single time-scale estimators and local chi-square ($\chi^2$) detectors to simultaneously opserve the states, share information, fuse the noise/attack-corrupted data locally, and detect possible anomalies in their own observations. While this scheme is applicable to a wide variety of systems associated with full-rank (invertible) matrices, we discuss  it within the context of distributed inference in social networks.
The proposed technique outperforms  existing results in the sense that: (i)  we  consider Gaussian noise with no simplifying upper-bound assumption on the support; (ii)
all existing $\chi^2$-based techniques are centralized while our proposed technique  is distributed, where the sensors \textit{locally} detect attacks, with no central coordinator, using specific probabilistic thresholds; and (iii) no local-observability assumption at a sensor is made, which makes our method feasible for large-scale social networks. Moreover, we consider a Linear Matrix Inequalities (LMI) approach to design block-diagonal gain (estimator) matrices under appropriate constraints for isolating the attacks. 
\end{abstract}

\begin{keywords}
Attack detection and isolation, Kronecker-product network, distributed estimation, $\chi^2$-test.
\end{keywords}

%
%
%
%

\section{Introduction} 
\label{sec_intro}

The unprecedented large size of social networks mandates distributed sensing, inference, and detection~\cite{SNAM20,block2020social,wai2016active,khan2014collaborative,pequito2014minimum,isj2020,jstsp14}, where the information is collected and processed locally while meeting certain security concerns. Recent distributed estimation protocols~\cite{isj2020,jstsp14,kar_cons_innov,mitra2018distributed,duan2020distributed,He_secure} are prone to faults/attacks that may result in inaccurate state estimates. Different attack detection and FDI (fault detection and isolation) strategies are thus proposed in the literature, ranging in applications from biological modeling \cite{mansouri2017improved} to smart-grid monitoring \cite{karimipour2019deep,ozay2015machine}, and from centralized approaches \cite{brumback1987chi,fawzi2014secure,kim2018detection,chong2015observability,tunga2018tuning,Gou_chi_square} to more recent distributed methods \cite{TCNS20,deghat2019detection,satchidanandan2016dynamic}. Among the centralized solutions, deterministic FDI and attack detection methods design decision thresholds based on the upper-bound on the noise support  \cite{kim2018detection,chong2015observability}, while, in contrast, probabilistic $\chi^2$-test with no such assumption on the noise is proposed in \cite{brumback1987chi} and further developed in \cite{tunga2018tuning,Gou_chi_square}. 
Among the distributed strategies, \cite{satchidanandan2016dynamic}  requires injecting a
watermarking input signal conceding to a loss in the control/estimation performance, which is not applicable to \textit{autonomous} systems (such as the social network model in this paper). In  order to close this gap, this paper aims at developing a  technique for distributed inference of autonomous (social) systems  while simultaneously detecting and isolating adversarial attacks \textit{locally} with no central coordinator.

The main contributions of this paper are as follows. (i) This work considers a windowed $\chi^2$ benchmark to \textit{locally} design probabilistic decision thresholds based on certain false alarm rates (FARs). This is in contrast to existing \textit{deterministic} thresholds assuming certain upper-bound  on the noise support \cite{kim2018detection,chong2015observability}, which results in faulty outcome when the noise upper-bound is considerably larger than the attack/fault magnitude. (ii)  This work extends the recent \textit{centralized} $\chi^2$ detectors \cite{tunga2018tuning,Gou_chi_square} to \textit{distributed}  ones, where the sensors are widespread over a large social network and, thus, the centralized solutions are infeasible/undesirable due to heavy communication loads or inability for parallel processing. In this direction, the notion of \textit{Kronecker-product network} \cite{TSIPN20} is used to perceive (structural) observability of the composite social/sensor network, which allows to find minimal connectivity requirement on the sensor network for distributed estimation/detection. (iii) Our distributed technique, as in \cite{TCNS20,deghat2019detection}, does not require local-observability at every sensor. However, unlike \textit{fixed} biasing faults/attacks on sensor outputs in \cite{TCNS20,deghat2019detection}, this work extends the results to general  anomalies in the form of a \textit{random} variable. In particular, we adopt the notion of \textit{distance measure} \cite{tunga2018tuning}, a scalar variable to compare the residual variance in presence and absence of attacks. 


\section{Problem Formulation} 
\label{sec_prob}
We consider the interaction of individuals in a social network as a linear-structure-invariant (LSI) autonomous model \cite{khan2014collaborative,pequito2014minimum,jstsp14,isj2020},
\begin{eqnarray}\label{eq_sys1}
\mb{x}_{k+1} = A\mb{x}_k + \nu_k,\qquad k\geq0,
\end{eqnarray}
where $k$ is the time-step, $A$ is the social system matrix associated with social digraph $\mc{G}$, $\nu_k \sim \mc{N}(0,Q)$ is  additive i.i.d noise vector, and  vector ${\mb{x}_k=\left[ {x}_{k}^1 ,\dots,{x}_{k}^n \right]^\top \in\mathbb{R}^n}$ represents the global social state. 
Note that $n$ is the size of social network and $\mb{x}_{k}^i$ represents the $i$'th individual's social state, e.g., opinion, rumor, or attitude \cite{SNAM20,block2020social,wai2016active,khan2014collaborative,isj2020,jstsp14,pequito2014minimum}. The state $\mb{x}_k^i$ of individual $i$ at time $k$ is a weighted average of the states $\mb{x}_{k-1}^{j}$ of its in-neighbors in $\mc{G}$  and its \textit{own previous state} $\mb{x}_{k-1}^{i}$. This is well-justified for opinion-dynamics in social systems, and particularly implies that  matrix $A$ is (structurally) full-rank \cite{jstsp14}.   
Consider
$N$ social sensors (agents or information-gatherers \cite{pequito2014minimum}) sensing the state of some individuals as,
\begin{eqnarray} \label{eq_H_i}
{y}_k^i = H_i\mb{x}_k + \tau_k^i + \eta_k^i,
\end{eqnarray}
with $H_i$ as the measurement matrix, $\tau_k^i$ as possible attack and $\eta_k^i \sim \mc{N}(0,R_i)$ as  Gaussian noise at sensor $i$ at time $k$. Define $R=\text{diag}[R_i]$ as the covariance matrix of the i.i.d noise vector $\eta_k$. Throughout this paper, without loss of generality,  we assume every sensor observes one  state variable, i.e., ${y}_k^i \in \mathbb{R}$.
Further, as in similar works \cite{brumback1987chi,tunga2018tuning,Gou_chi_square}, we assume the system and measurement noise covariance ($Q$ and $R$) are known. 
Then, sensors share their information over a sensor network $\mc{G}_N$. Clearly, system $A$ is not locally observable to any sensor, but globally observable to all sensors. The condition on $(A,H)$-observability is given in the following lemma.  
\begin{lem}\label{lem_scc}
\cite{jstsp14} Given a social network $\mc{G}$ (with structurally full-rank adjacency matrix $A$), if at least  one social state is sensed  in every strongly-connected-component (SCC) in $\mc{G}$, then, the pair $(A,H)$ is (structurally) observable. 
\end{lem}

Given (social) system \eqref{eq_sys1} and state observations \eqref{eq_H_i} satisfying Lemma~\ref{lem_scc}, we aim to design a distributed iterative procedure to simultaneously estimate the (social) state $\mb{x}_k^i$ while detecting adversary attacks at (social) sensors. The attack by the  adversary is modeled as an additive random term $\tau_k^i$ at sensor $i$ in \eqref{eq_H_i}.
The proposed distributed estimation makes the entire  system observable to every sensor, and the attack-detection technique enables each sensor to locally detect anomalies in its observation with a certain FAR (false-alarm rate).    

\section{Main Algorithm} \label{sec_main}
We consider a modified version of the single time-scale distributed estimator in \cite{jstsp14} with one step of averaging on \textit{a-priori} estimates (similar to \textit{DeGroot consensus model} \cite{kar_cons_innov}) and one step of measurement update (also known as \textit{innovation} \cite{kar_cons_innov}), 
\begin{eqnarray}\label{eq_p}
\widehat{\mb{x}}^i_{k|k-1} &=& \sum_{j\in\mathcal{N}(i)} w_{ij}A\widehat{\mb{x}}^j_{k-1|k-1}, \\ \label{eq_m}
\widehat{\mb{x}}^i_{k|k} &=&\widehat{\mb{x}}^i_{k|k-1} + K_i H_i^\top \left(y^i_k-H_i\widehat{\mb{x}}^i_{k|k-1}\right).
\end{eqnarray}
with \textit{stochastic matrix} $W = \{w_{ij}\}$ as the adjacency  matrix of the sensor network $\mc{G}_N$ representing  the fusion weights among the sensors, $K_i$ as the local gain matrix at agent~$i$, and $\widehat{\mb{x}}^i_{k|k-1}$ and $\widehat{\mb{x}}^i_{k|k}$ as the state estimate at time $k$  given all the information of agent~$i$ and its in-neighbors $\mathcal{N}(i)$, respectively,  at time~$k-1$ and $k$. In contrast to double time-scale estimators/observers \cite{He_secure} with many consensus iterations between every two consecutive time-steps $k-1$ and $k$ of social dynamics~\eqref{eq_sys1}, the estimator \eqref{eq_p}-\eqref{eq_m} performs one iteration of information fusion between steps $k-1$ and $k$, which is more efficient in terms of computation/communication loads. 

Define the estimation error at agent $i$ as $\mb{e}_{k}^i = \mb{x}_{k} - \widehat{\mb{x}}^i_{k|k}$ and the error vector $\mb{e}_k=\left[ (\mb{e}_{k}^1)^\top,\dots,(\mb{e}_{k}^n)^\top \right]^\top$. Following similar procedure as in~\cite{isj2020}, the error dynamics is as follows,
\begin{eqnarray}\label{eq_err1}
\mb{e}_{k} = (W\otimes A - KD_H(W\otimes A))\mb{e}_{k-1} +
\mb{q}_k,
\end{eqnarray}
with $D_H=\text{diag}[H_i^\top H_i]$, $K=\text{diag}[K_i]$ as the feedback gain matrix, and $\mb{q}_k$ as the collective vector of noise-related terms $\mb{q}_k=\left[ (\mb{q}_{k}^1)^\top,\dots,(\mb{q}_{k}^n)^\top \right]^\top$ as,
\begin{align} 
\mb{q}_k^i =& \mb{\nu}_{k-1}-K_i \Bigl(H_i^\top \mb{\eta}^i_{k} +
H_i^\top \mb{\tau}^i_{k}+ H_i^\top H_i\mb{\nu}_{k-1}\Bigr),
\label{eq_q}
\\ \nonumber
\mb{q}_k
=&  \mathbf{1}_N \otimes \mb{\nu}_{k-1}- K D_H(\mathbf{1}_N \otimes \mb{\nu}_{k-1}) \\ &- K\overline{D}_H\mb{\eta}_{k} -K\overline{D}_H\mb{\tau}_{k},
\label{eq_q2}
\end{align}
with $\mathbf{1}_N$ as the vector of $1$'s of size $N$ and $\overline{D}_H=\mbox{diag}[H_i^\top]$.
Following Kalman theory, for bounded steady-state estimation error,  $(W \otimes A, D_H)$ needs to be observable, characterizing the \textit{distributed observability} condition for network of estimators/observers \cite{globalsip14}. Using structured system theory, this condition can be investigated via graph theoretic notions. In this direction, the associated network to $W \otimes A$ is a  Kronecker-product network, whose observability condition relies on the structure of both $\mc{G}$ and $\mc{G}_N$. Given the social network $\mc{G}$, the conditions on the sensor network $\mc{G}_N$ to satisfy distributed observability follows the recent results on \textit{composite-network theory} and network observability  discussed in \cite{TSIPN20}, which is summarized in the following lemma.

\begin{lem}\label{lem_kron}
\cite{TSIPN20} Given $(A,H)$-observability via Lemma~\ref{lem_scc}, minimal sufficient condition for $(W \otimes A, D_H)$-observability is that matrix $W$ be irreducible, i.e., the network $\mc{G}_N$ be strongly-connected (SC).
\end{lem}

For an observable pair $(W \otimes A, D_H)$, the feedback gain matrix $K$ can be designed to stabilize the error dynamics~\eqref{eq_err1}. Mathematically, for $\overline{A}=W\otimes A - KD_H(W\otimes A)$, we need to design $K$ such that $\rho(\overline{A})<1$ (Schur stability of error dynamics \eqref{eq_err1}) for general  social systems with $\rho(A)>1$  with $\rho(\cdot)$ as the spectral radius. As mentioned before, for distributed case, $K$ needs to be further block-diagonal such that each sensor only uses local information in its own neighborhood. The iterative LMI-based algorithm to design such block-diagonal gain $K$ is given in \cite{usman_cdc:11}. In attack-free scenario, the distributed estimator/observer \eqref{eq_p}-\eqref{eq_m} with proper gain $K$  ensures tracking the global social state with bounded steady-state error as discussed in \cite{isj2020,jstsp14}. Next, in this section, we further study the performance of the proposed protocol in the presence of non-zero random attack signals.
Define  $\widehat{y}_{k}^i = H_i \widehat{\mb{x}}_{k|k}^i$ as the estimated output at sensor $i$ at time $k$. To detect possible attacks, each sensor calculates its \textit{residual} as the difference of its original output and the estimated one,
\begin{align}\label{eq_residual}
r_k^i =& {y}_k^i-\widehat{y}_{k}^i={y}_k^i-H_i \widehat{\mb{x}}_{k|k}^i= H_i\mb{e}_{k}^i+\mb{\eta}^i_k+\mb{\tau}^i_k .
\end{align}
Having $\rho(\overline{A})<1$,  the steady-state error in \eqref{eq_err1} only relies on the term $\mb{q}_k^i$  defined in \eqref{eq_q} as,  
\begin{align}\nonumber
H_i\mb{q}_k^i &= H_i\mb{\nu}_{k-1}-H_iK_iH_i^\top \mb{\eta}^i_{k} \\
&+
H_iK_iH_i^\top \mb{\tau}^i_{k}+ H_iK_iH_i^\top H_i\mb{\nu}_{k-1}.
\label{eq_residual2}
\end{align}
From \eqref{eq_residual} and \eqref{eq_residual2}, it is clear that for  $\mb{\tau}^i_{k}\neq 0$,  only the residual $r_k^i $ at sensor $i$ is biased with no effect on the residual of other sensors $j\neq i$. This allows to \textit{isolate} the attacked sensor as  $r_k^i$  only depends on $\mb{\tau}^i_{k}$ and not on $\mb{\tau}^j_{k}$'s. To detect a possible attack at agent $i$ via residual $r_k^i$, the non-zero term $\mb{\tau}^i_k-H_iK_iH_i^\top \mb{\tau}_k^i$ needs to be sufficiently larger than other noise terms. This ensures that the residual in attacked case is large enough to be distinguished from the noise terms in attack-free case.   
This is done by adding the constraint $|1-H_iK_iH_i^\top|>C$ (with $C$ as a pre-selected lower-bound) to the proposed LMI in \cite{usman_cdc:11}. To the best of our knowledge, no other analytical solution for such constrained design is given in the literature.
Clearly, the detecting probability of an  attack depends on the magnitude of $\mb{\tau}^i_k$, which justifies the \textit{probabilistic threshold design}. 
In this direction, we consider a distributed probability-based $\chi^2$-test which outperforms the deterministic fault/attack detection methods as  it considers noise of \textit{unbounded support}. In this case, instead of a deterministic threshold with $0$ (no attack) or $1$ (attack detected) outcome, different probabilistic thresholds (with different sensitivities) are defined each assigned with an FAR. In fact, higher \textit{residual-to-noise ratio} (RNR) stimulates the threshold with lower FAR. 
In this direction, first the covariance of  error $\mb{e}_k$ and (attack-free) residuals need to be calculated, which are tied with the noise covariance $Q$ and $R$. Let $\Xi_k = \mathbb{E}(\mb{e}_k\mb{e}_k^{\top})$ and $\Sigma =\mathbb{E}(\mb{q}_k\mb{q}^\top_k) $. Then, from \eqref{eq_err1}, 
\begin{eqnarray} 
   \Xi_{k} 
   &=& \overline{A}^k \Xi_{0}(\overline{A}^k)^\top + \sum_{j=1}^{k-1}  \overline{A}^{j}\Sigma(\overline{A}^{j})^\top + \Sigma.
   \label{eq_xi}
\end{eqnarray}
Knowing that $\rho(\overline{A})<1$, the first term in \eqref{eq_xi} goes to zero. Therefore, it can be proved from \cite{khan2014collaborative} that   for $\Xi_{\infty} = \lim_{k \rightarrow \infty} \Xi_{k} $,
\begin{eqnarray} \label{eq_pinfty1}
\|\Xi_{\infty}\|_2 = \|\sum_{j=1}^{\infty}  \overline{A}^{j}\Sigma(\overline{A}^j)^\top+\Sigma\|_2 \leq \frac{\|\Sigma\|_2}{1-b^2},
\end{eqnarray}
with $b = \|\overline{A}\|_2<1 $. For  attack-free case ($\mb{\tau}_{k}=\mb{0}_N$ in \eqref{eq_q}),
\begin{align} \nonumber
 \mb{q}_k\mb{q}^\top_k &= (I_{Nn}- K D_H)(\mathbf{1}_{NN} \otimes \nu_{k-1}\nu_{k-1}^\top)(I_{Nn}- K D_H)^\top\\ &+ (K\overline{D}_C) \eta_k\eta_k^\top (K\overline{D}_H)^\top ,
\end{align}
where $\mathbf{1}_{NN}$ is the $1$'s matrix of size $N$. Applying the $\mathbb{E}(\cdot)$ and $2$-norm operators,
\begin{align} \nonumber
\|\Sigma\|_2  &= \|(I_{Nn}- K D_H)(\mathbf{1}_{NN} \otimes Q)(I_{Nn}- K D_H)^\top\|_2\\ &+ \|(K\overline{D}_H) R (K\overline{D}_H)^\top\|_2 .
\end{align}
Then, the upper-bound on $\|\Sigma\|_2$ is,
\begin{eqnarray} \nonumber
\|\Sigma\|_2 \leq \|I_{Nn}- K D_H\|_2^2 N \|Q\|_2 + \|K\|_2^2 \|\overline{R}\|_2,
\end{eqnarray}
with $\overline{R}=\mbox{diag}[H_i^\top R_iH_i]$.
Then, using \eqref{eq_pinfty1},
\begin{eqnarray} \label{eq_pinfty}
\frac{\|\Xi_{\infty}\|_2}{N}  \leq \frac{a_1N\|Q\|_2+a_2 a_3 \|R\|_2}{N(1-b^2)}=\Phi , 
\end{eqnarray}
where $\|I_{Nn}- K D_H\|_2^2=a_1$, $\|K\|_2^2=a_2$, and $\|\overline{R}\|_2=a_3 \|R\|_2$. Note that in \eqref{eq_pinfty} the error covariance is scaled  by the number of sensors $N$. From \eqref{eq_pinfty},  assuming no attack is present ($\mb{\tau}_{k}^i=0$), a conservative approximation for error variance at sensor $i$ is $\mathbb{E}((\mb{e}^i_k) (\mb{e}_{k}^i)^\top) = \Phi$. Then, following the discussion in \cite{tunga2018tuning}, the residual $r^i_k$  in \eqref{eq_residual}  can be assumed as a zero-mean Gaussian variable with maximum variance ${\Lambda_i = \mathbb{E}((r^i_k) (r_{k}^i) ^\top)=H_i^\top \Phi H_i+R_i}$, i.e., $r_{k}^i \sim \mc{N}(0, \Lambda_i)$. Define,
\begin{equation} \label{eq_z}
    z_{k}^i  = \frac{(r_{k}^i)^2}{\Lambda_i},~ v_{k}^i = \sum_{t=k-T+1}^k z_{t}^i, 
\end{equation}
with $T$ as the length of the \textit{sliding window}\footnote{In general, each agent can consider a different length for the horizon T.}.
It is known that, for a Gaussian variable $r_{k}^i$, scalars $z_{k}^i$ and $v_{k}^i$ follow $\chi_1^2$-distribution with degree $1$ and $T$ respectively ($\mathbb{E}[z_{k}^i]=1$ and $\mathbb{E}[v_{k}^i]=T$) \cite{chi_book}. In fact, these 
so-called \textit{distance measures} $z_{k}^i$ and $v_{k}^i$ give an estimate of  variance of $r_{k}^i$ relative to the attack-free variance $\Lambda_i$ \cite{tunga2018tuning}, and are known to outperform simple detectors  comparing \textit{absolute residual} to a threshold as in \cite{TCNS20,deghat2019detection}. 
Next, we determine the \textit{decision threshold}  $\theta$ on $v_{k}^i$ based on a pre-specified FAR $p$. It can be shown that $p=1-F(\theta)$ where $F(\cdot)$ is the cumulative distribution function (CDF) of $\chi_1^2$-distribution. Then,
\begin{equation} \label{eq_theta}
\theta = 2\Gamma^{-1}(1-p,\frac{T}{2}),
\end{equation}
with $\Gamma^{-1}(\cdot,\cdot)$ as the \textit{inverse regularized lower incomplete gamma function} \cite{chi_book}. Using \eqref{eq_theta}, our attack detection logic at each sensor $i$ is as follows,
\begin{equation}
  \text{If}~ \left\{
  \begin{array}{@{}l}
     v_{k}^i\geq \theta \\
     v_{k}^i<\theta 
  \end{array}\right. ~\text{Then}~\left\{
  \begin{array}{@{}l}
     \mc{H}^i_1: \text{Attack Detected} \\
     \mc{H}^i_0: \text{No Attack} 
  \end{array}\right. 
\end{equation} 
It should be noted that the existing $\chi^2$-based attack detection scenarios in literature are all centralized \cite{brumback1987chi,tunga2018tuning,Gou_chi_square} and in this work, using distributed estimation, we enable detection of attacks \textit{locally} at every sensor with no need of a central unit. 
We summarize our proposed simultaneous distributed estimation and attack detection technique  in Algorithm~1.
\begin{algorithm} \label{alg_logic}
	\textbf{Given:} System matrix $A$, Network $\mc{G}_N$, Fusion matrix $W$, Measurements ${y}_k$, Measurement matrix $H$, System/Measurement noise covariance $Q$/$R$, false-alarm probability (FAR) $p$, sliding window $T$ 
	
	Choose block-diagonal gain $K$ via LMI in \cite{usman_cdc:11}\;
	Find $\widehat{\mb{x}}^i_{k|k}$ at every sensor $i$ via \eqref{eq_p}-\eqref{eq_m}\;
	Find $\Lambda_i$ based on $R$, $Q$, and \eqref{eq_pinfty}\;
	Find residuals $r_k^i$ at every sensor $i$ via \eqref{eq_residual}\;	
	Find $z_k^i$ and $v_{k}^i$ at every sensor $i$ via \eqref{eq_z} \;
	Define threshold $\theta$ based on $p$ and $T$ via \eqref{eq_theta}\;
	If $v_k^i\geq \theta$ return $\mc{H}^i_1$: {Attack Detected} with FAR $p$\;
	If $v_k^i<\theta$ return $\mc{H}^i_0$: {No Attack}\;
	\textbf{Return:} Hypothesis $\mc{H}^i_0$ or $\mc{H}^i_1$ for $i = \{1,...,N\}$.
	\caption{\small{Proposed iterative methodology.}}
\end{algorithm}

Note that after detecting a malicious attack with low FAR, the strategy in \cite{SPL17} can be adopted to remove unreliable data and replace the compromised sensor with its \textit{observationally equivalent} counterpart to regain distributed observability. 

\section{Simulation Results} 
\label{sec_sim}
\begin{figure}[t]	
	\centering
	\includegraphics[width=1.6in]{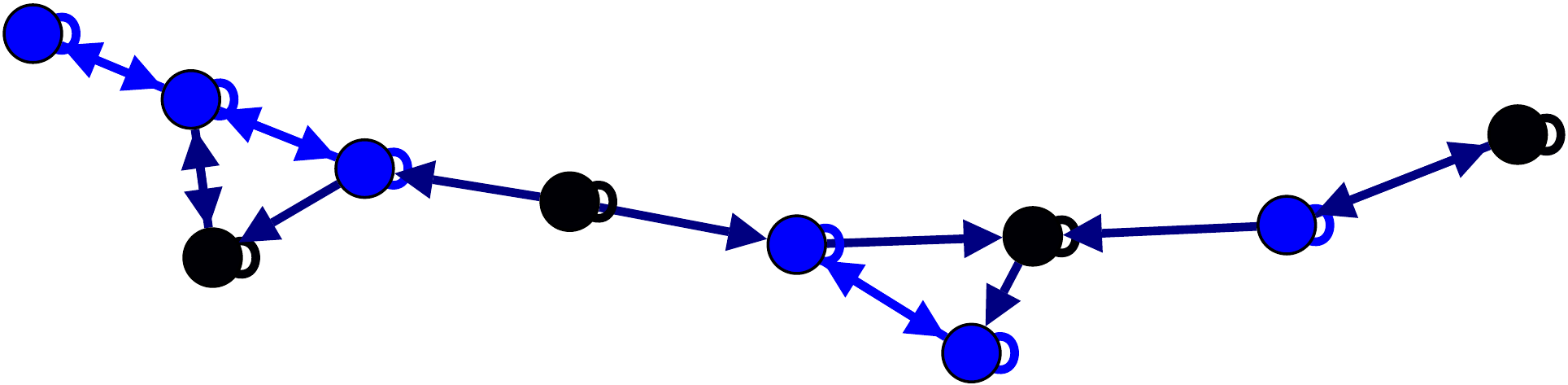}
	\caption{The small  social network  $\mc{G}$ considered for simulation.  The black state nodes are observed by the sensors (satisfying Lemma~\ref{lem_scc}). }\label{fig_net}
\end{figure}

\begin{figure}[t]	
	\centering
	\includegraphics[width=1.5in]{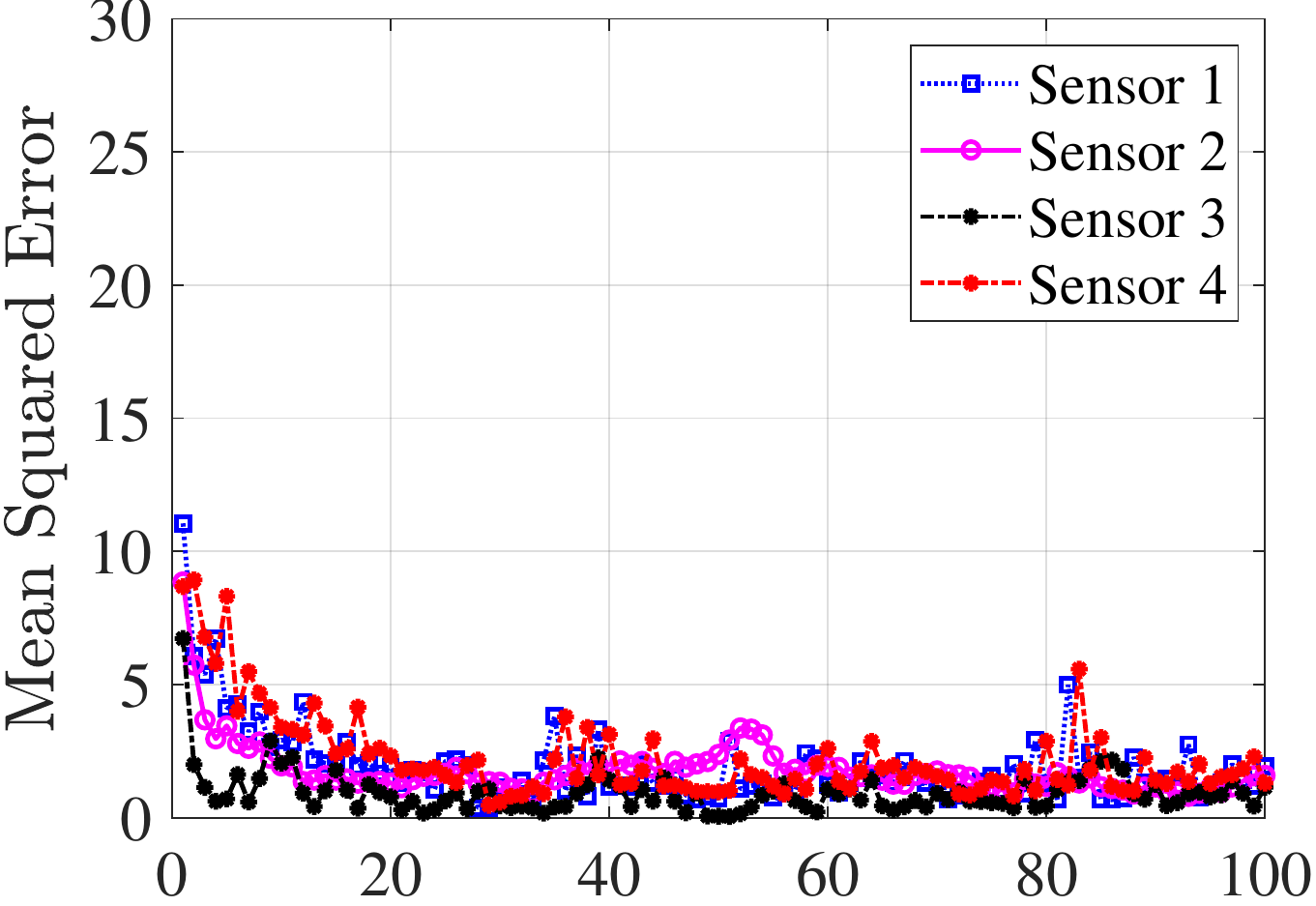}
	\includegraphics[width=1.5in]{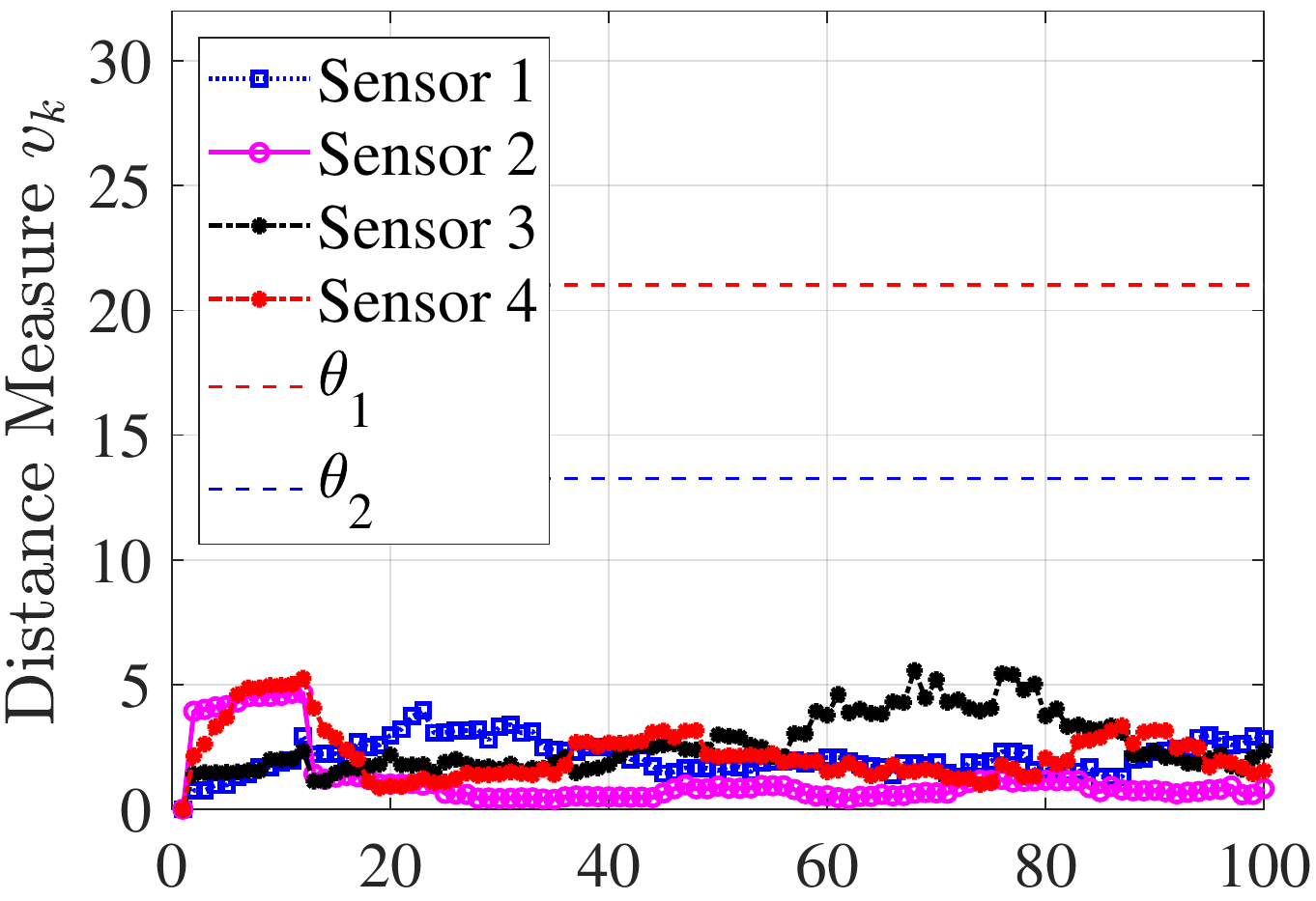}
	\includegraphics[width=1.5in]{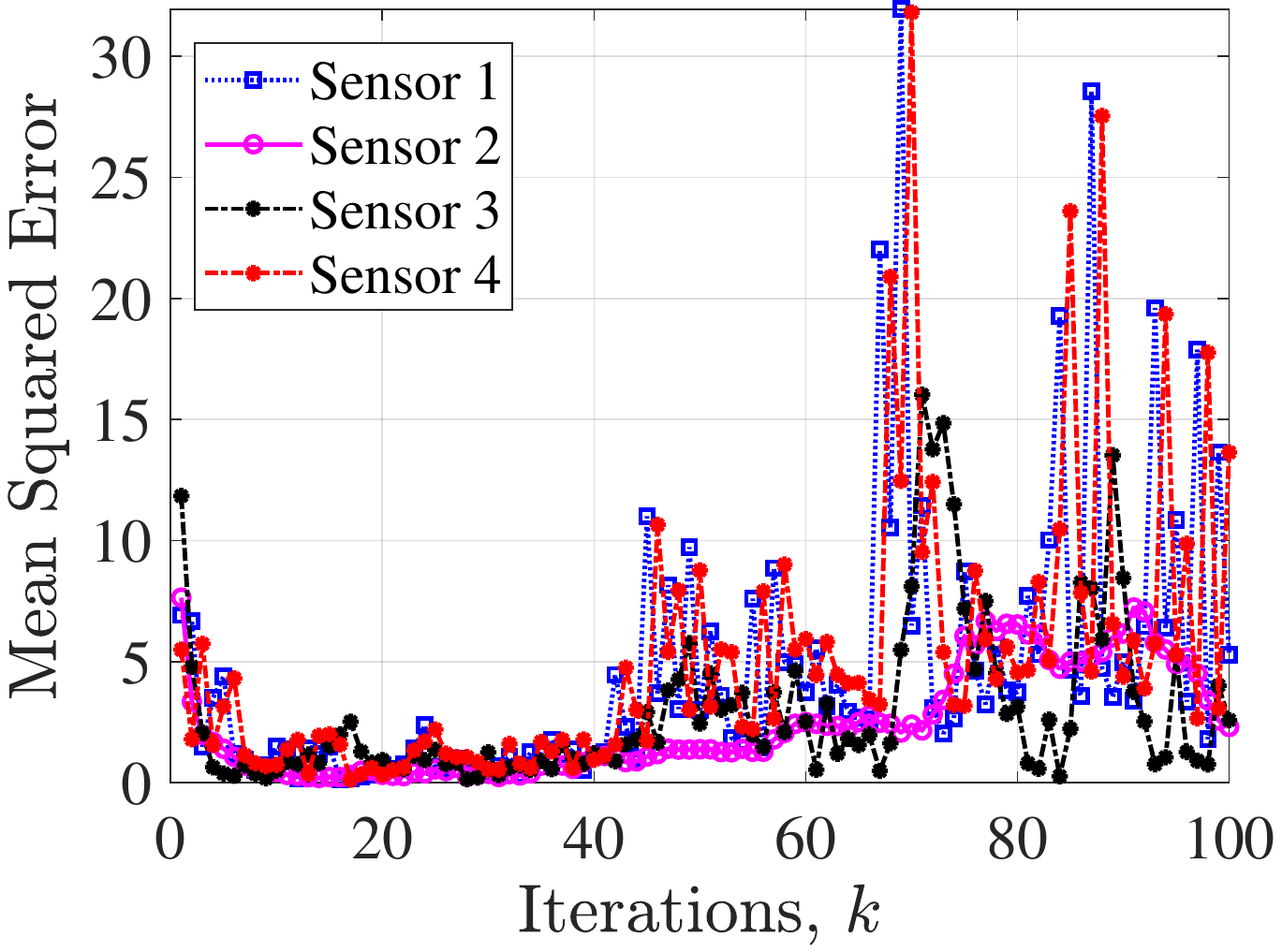}
	\includegraphics[width=1.5in]{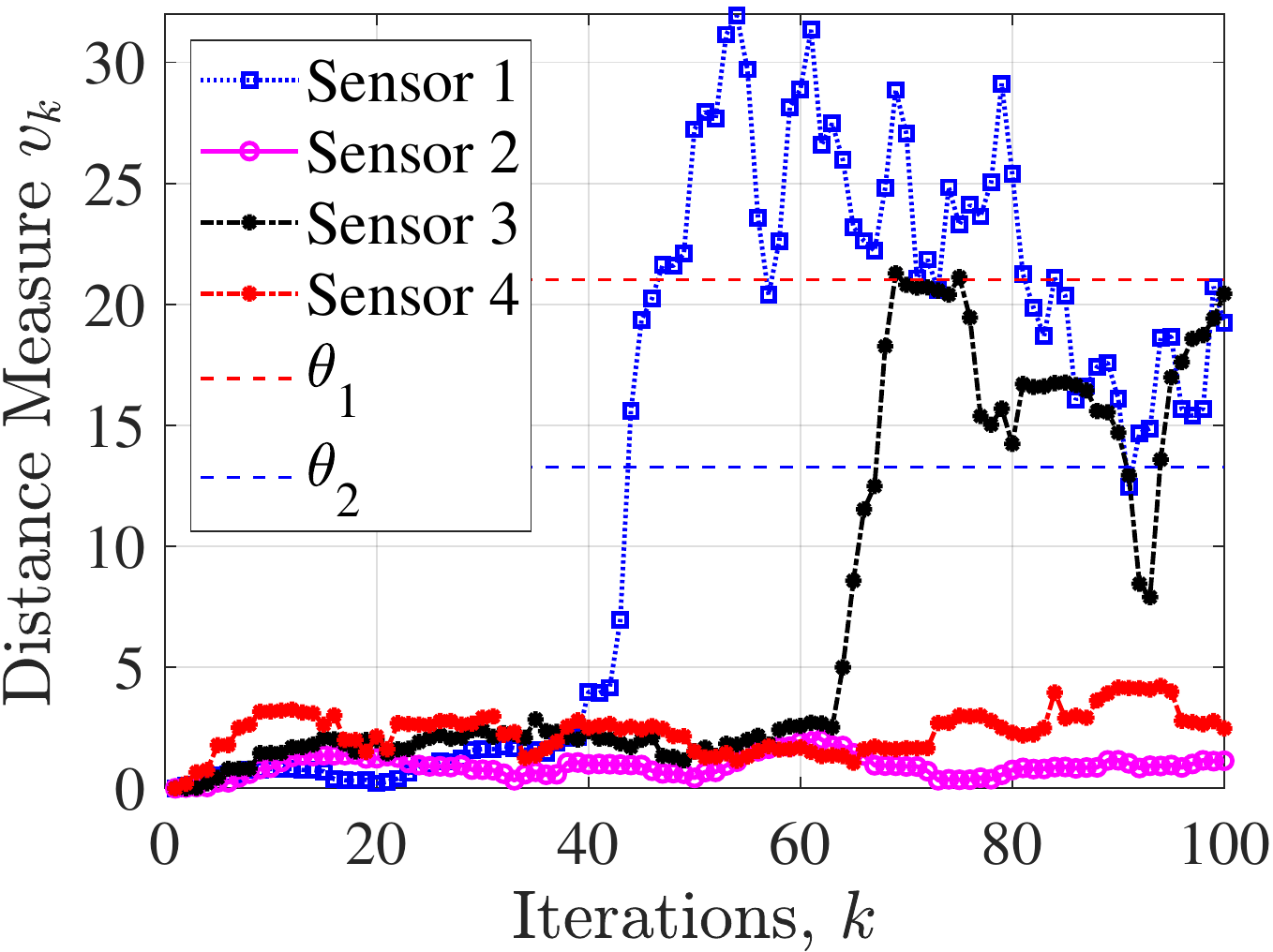}	
	\caption{(top) No attack: mean squared estimation error at all sensors are steady-state stable. (bottom) Attack at sensors $1$ and $3$: the non-zero attacks add bias to the estimation error at all sensors. Distance measures $v_k^1$ and $v_k^3$ exceeding $\theta_2$ reveal possible attacks at sensors $1$ and $3$ with FAR $p_2=35\%$, while $v_k^1$  exceeding $\theta_1$ implies lower FAR $p_1=5\%$ for attack at sensor $1$. 
	}\label{fig_err}
\end{figure}

We evaluate our theoretical results on an example social network $\mc{G}$ of $10$ state nodes with $4$ sensor observations shown in Fig.~\ref{fig_net}. The network $\mc{G}_N$ of $4$ sensors is considered as a cycle (satisfying Lemma~\ref{lem_kron}). 
The fixed non-zero entries of $A$ and $W$ are chosen randomly in  $(0,1.1]$.  Further, $\rho(A)=1.1$ implying a potentially unstable system, $\eta_k^i,\nu_k^i \sim \mc{N}(0,0.06)$, and non-zero entries of $H$ are set as $1$. Using MATLAB's \texttt{CVX}, the stabilizing  block-diagonal gain $K$ is designed via the iterative LMI in \cite{usman_cdc:11} subject to $|1-H_iK_iH_i^\top|>0.2$, which results in $\rho(\overline{A})=0.97$, $b=1.42$, $\Phi = 4.82$, and $\Lambda_i = 4.88$.  In attack-free case, each sensor is able to track the global social state $\mb{x}_k$ over time via protocol \eqref{eq_p}-\eqref{eq_m}. The time-evolution of mean squared errors $\frac{\|\mb{e}_k^i\|^2}{n}$ and distance measures at all sensors are shown in Fig.~\ref{fig_err}(top). Next, considering two non-zero attack sequences as $\tau_k^3 \sim \mc{N}(0.2,0.3)$ for $k\geq 60$ and $\tau_k^1 \sim \mc{N}(0,0.8)$ for $k\geq 40$, the distance measures $v_{k}^i$'s over a sliding window of  $T=12$-step length are shown in Fig.~\ref{fig_err}(bottom). The figure clearly shows that the attacks affect the estimation error at all sensors. Setting two FARs $p_1=5\%$, $p_2=35\%$, the associated decision thresholds are $\theta_1=21$, $\theta_2=13.3$ via \eqref{eq_theta}.
From the figure, the less conservative threshold $\theta_2$ reveals both attacks, while $\theta_1$ only detects one and the other one remains stealthy most of the times.

\section{Conclusion} 
\label{sec_con}
We proposed an algorithm for simultaneous estimation of states and attack detection over a distributed sensor network.
Using a windowed chi-square detector, every sensor is able to locally detect possible measurement anomalies causing the residuals to exceed an FAR-based threshold. 
As future research directions, the results in \cite{pequito2014minimum,spl18} can be adopted to optimally locate the sensing nodes and design the  network among the social  sensors to reduce cost. Additionally, adopting the pruning strategies in \cite{SNAM20,block2020social}, one can change the social network structure and, in turn, tune its observability and information flow  to improve estimation/detection properties.     

\bibliographystyle{IEEEbib}
\small{
\bibliography{bibliography}
}
\end{document}